\documentclass[11pt]{article}
\usepackage{times,epsfig,nicefrac,color,balance}
\usepackage{graphics,subfigure, cuted}

\usepackage{cite,calc}

\usepackage{amsmath,amsthm,mathtools}
\usepackage{amssymb}
\usepackage{mathrsfs}

\usepackage{multirow}
\usepackage{array}

\usepackage{amsfonts}
\usepackage{graphicx}
\usepackage[linesnumbered]{algorithm2e}

\usepackage{color, verbatim}
\usepackage[usenames,dvipsnames]{xcolor}
\usepackage{url}
\usepackage{cleveref}

\usepackage{epsfig}
\usepackage{subfig}
\usepackage{epstopdf}
\usepackage{xargs}

\newcommand\nc\newcommand
\nc{\bb}[1]{\mathbb{#1}}
\renewcommand{\cal}[1]{\mathcal{#1}}
\renewcommand{\bf}[1]{\mathbf{#1}}

\DeclarePairedDelimiter{\floor}{\lfloor}{\rfloor}
\DeclarePairedDelimiter{\set}{\lbrace}{\rbrace}
\DeclarePairedDelimiter{\br}{\lparen}{\rparen}
\DeclarePairedDelimiter{\brac}{\lbrack}{\rbrack}
\DeclarePairedDelimiter{\abs}{\lvert}{\rvert}

\nc\bfa{{\bf{a}}}\nc\bfA{{\boldsymbol A}}\nc\cA{{\cal A}} \nc\fA[1]{A\br*{#1}} \nc\fa[1]{a\br*{#1}}  \nc\rmA{\mathrm{A}} \nc\rma{\mathrm{a}}
\nc\bfb{{\bf{b}}}\nc\bfB{{\boldsymbol B}}\nc\cB{{\cal B}} \nc\fB[1]{B\br*{#1}} \nc\fb[1]{b\br*{#1}}  \nc\rmB{\mathrm{B}} \nc\rmb{\mathrm{b}}
\nc\bfc{{\bf{c}}}\nc\bfC{{\boldsymbol C}}\nc\cC{{\cal C}} \nc\fC[1]{C\br*{#1}} \nc\fc[1]{c\br*{#1}}  \nc\rmC{\mathrm{C}} \nc\rmc{\mathrm{c}}
\nc\bfd{{\bf{d}}}\nc\bfD{{\boldsymbol D}}\nc\cD{{\cal D}} \nc\fD[1]{D\br*{#1}} \nc\fd[1]{d\br*{#1}}  \nc\rmD{\mathrm{D}} \nc\rmd{\mathrm{d}}
\nc\bfe{{\bf{e}}}\nc\bfE{{\boldsymbol E}}\nc\cE{{\cal E}} \nc\fE[1]{E\br*{#1}} \nc\fe[1]{e\br*{#1}}  \nc\rmE{\mathrm{E}} \nc\rme{\mathrm{e}}
\nc\bff{{\bf{f}}}\nc\bfF{{\boldsymbol F}}\nc\cF{{\cal F}} \nc\fF[1]{F\br*{#1}} \nc\ff[1]{f\br*{#1}}  \nc\rmF{\mathrm{F}} \nc\rmf{\mathrm{f}}
\nc\bfg{{\bf{g}}}\nc\bfG{{\boldsymbol G}}\nc\cG{{\cal G}} \nc\fG[1]{G\br*{#1}} \nc\fg[1]{g\br*{#1}}  \nc\rmG{\mathrm{G}} \nc\rmg{\mathrm{g}}
\nc\bfh{{\bf{h}}}\nc\bfH{{\boldsymbol H}}\nc\cH{{\cal H}} \nc\fH[1]{H\br*{#1}} \nc\fh[1]{h\br*{#1}}  \nc\rmH{\mathrm{H}} \nc\rmh{\mathrm{h}}
\nc\bfi{{\bf{i}}}\nc\bfI{{\boldsymbol I}}\nc\cI{{\cal I}} \nc\fI[1]{I\br*{#1}} \nc\rmI{\mathrm{I}} \nc\rmi{\mathrm{i}}
\nc\bfj{{\bf{j}}}\nc\bfJ{{\boldsymbol J}}\nc\cJ{{\cal J}} \nc\fJ[1]{J\br*{#1}} \nc\fj[1]{j\br*{#1}} \nc\rmJ{\mathrm{J}} \nc\rmj{\mathrm{j}}
\nc\bfk{{\bf{k}}}\nc\bfK{{\boldsymbol K}}\nc\cK{{\cal K}} \nc\fK[1]{K\br*{#1}} \nc\fk[1]{k\br*{#1}} \nc\rmK{\mathrm{K}} \nc\rmk{\mathrm{k}}
\nc\bfl{{\bf{l}}}\nc\bfL{{\boldsymbol L}}\nc\cL{{\cal L}} \nc\fL[1]{L\br*{#1}} \nc\fl[1]{l\br*{#1}} \nc\rmL{\mathrm{L}} \nc\rml{\mathrm{l}}
\nc\bfm{{\bf{m}}}\nc\bfM{{\boldsymbol M}}\nc\cM{{\cal M}} \nc\fM[1]{M\br*{#1}} \nc\fm[1]{m\br*{#1}} \nc\rmM{\mathrm{M}} \nc\rmm{\mathrm{m}}
\nc\bfn{{\bf{n}}}\nc\bfN{{\boldsymbol N}}\nc\cN{{\cal N}} \nc\fN[1]{N\br*{#1}} \nc\fn[1]{n\br*{#1}} \nc\rmN{\mathrm{N}} \nc\rmn{\mathrm{n}}
\nc\bfo{{\bf{o}}}\nc\bfO{{\boldsymbol O}}\nc\cO{{\cal O}} \nc\fO[1]{O\br*{#1}} \nc\fo[1]{o\br*{#1}} \nc\rmO{\mathrm{O}} \nc\rmo{\mathrm{o}}
\nc\bfp{{\bf{p}}}\nc\bfP{{\boldsymbol P}}\nc\cP{{\cal P}} \nc\fP[1]{P\br*{#1}} \nc\fp[1]{p\br*{#1}} \nc\rmP{\mathrm{P}} \nc\rmp{\mathrm{p}}
\nc\bfq{{\bf{q}}}\nc\bfQ{{\boldsymbol Q}}\nc\cQ{{\cal Q}} \nc\fQ[1]{Q\br*{#1}} \nc\fq[1]{q\br*{#1}} \nc\rmQ{\mathrm{Q}} \nc\rmq{\mathrm{q}}
\nc\bfr{{\bf{r}}}\nc\bfR{{\boldsymbol R}}\nc\cR{{\cal R}} \nc\fR[1]{R\br*{#1}} \nc\fr[1]{r\br*{#1}} \nc\rmR{\mathrm{R}} \nc\rmr{\mathrm{r}}
\nc\bfs{{\bf{s}}}\nc\bfS{{\boldsymbol S}}\nc\cS{{\cal S}} \nc\fS[1]{S\br*{#1}} \nc\fs[1]{s\br*{#1}} \nc\rmS{\mathrm{S}} \nc\rms{\mathrm{s}}
\nc\bft{{\bf{t}}}\nc\bfT{{\boldsymbol T}}\nc\cT{{\cal T}} \nc\fT[1]{T\br*{#1}} \nc\ft[1]{t\br*{#1}} \nc\rmT{\mathrm{T}} \nc\rmt{\mathrm{t}}
\nc\bfu{{\bf{u}}}\nc\bfU{{\boldsymbol U}}\nc\cU{{\cal U}} \nc\fU[1]{U\br*{#1}} \nc\fu[1]{u\br*{#1}} \nc\rmU{\mathrm{U}} \nc\rmu{\mathrm{u}}
\nc\bfv{{\bf{v}}}\nc\bfV{{\boldsymbol V}}\nc\cV{{\cal V}} \nc\fV[1]{V\br*{#1}} \nc\fv[1]{v\br*{#1}} \nc\rmV{\mathrm{V}} \nc\rmv{\mathrm{v}}
\nc\bfw{{\bf{w}}}\nc\bfW{{\boldsymbol W}}\nc\cW{{\cal W}} \nc\fW[1]{W\br*{#1}} \nc\fw[1]{w\br*{#1}} \nc\rmW{\mathrm{W}} \nc\rmw{\mathrm{w}}
\nc\bfx{{\bf{x}}}\nc\bfX{{\boldsymbol X}}\nc\cX{{\cal X}} \nc\fX[1]{X\br*{#1}} \nc\fx[1]{x\br*{#1}} \nc\rmX{\mathrm{X}} \nc\rmx{\mathrm{x}}
\nc\bfy{{\bf{y}}}\nc\bfY{{\boldsymbol Y}}\nc\cY{{\cal Y}} \nc\fY[1]{Y\br*{#1}} \nc\fy[1]{y\br*{#1}} \nc\rmY{\mathrm{Y}} \nc\rmy{\mathrm{y}}
\nc\bfz{{\bf{z}}}\nc\bfZ{{\boldsymbol Z}}\nc\cZ{{\cal Z}} \nc\fZ[1]{Z\br*{#1}} \nc\fz[1]{z\br*{#1}} \nc\rmZ{\mathrm{Z}} \nc\rmz{\mathrm{z}}

\DeclareMathOperator{\Log}{\log}
\DeclareMathOperator{\Exp}{\exp}

\DeclareMathOperator{\spn}{span}

\DeclareMathOperator{\wt}{wt}

\DeclareMathOperator{\prob}{\mathbb{P}}

\nc\defeq{\coloneqq}

\providecommand\given{}
\newcommand\SetSymbol[1][]{\nonscript\:#1\vert \allowbreak \nonscript\: \mathopen{}}

\DeclarePairedDelimiterX\Set[1]\{\}{\renewcommand\given{\SetSymbol[\delimsize]}#1}

\newcommand\F{{\mathbb F}}

\newcommand\integers{{\mathbb Z}}

\newtheorem{theorem}{Theorem}
\newtheorem{definition}{Definition}
\crefname{definition}{defn.}{defns}
\newtheorem{lemma}[theorem]{Lemma}

\newcommand\mmod{\hspace{-.6em}\mod}

\DeclareMathOperator{\Cl}{B_{\bb{T}}}

\nc\Clo{\Cl{\rho n}}
\nc\s{\floor*{\frac{r+1}{2}}}
\newcommandx*\vect[4][1=1, 3=n,4={}]{\begin{pmatrix}#2_{#1}#4\\ \vdots \\ #2_{#3}#4\end{pmatrix}}

\crefname{Appendix}{Appendix}{Appendices}

\begin{document}
\sloppy

\title{Bounds on the Rate of Linear Locally Repairable Codes over Small Alphabets}
\author{Abhishek Agarwal  and Arya Mazumdar
	\thanks{Abhishek Agarwal is with the Department of Electrical and Computer Engineering, University of Minnesota, Minneapolis, MN  55455. Arya Mazumdar is with the Computer Science Department of University of Massachusetts, Amherst, MA 01003, and was with the University of Minnesota. email: \texttt{abhiag@umn.edu, arya@cs.umass.edu}. Research supported by NSF grants CCF 1318093, CCF 1642658, CCF 1453121 and CCF 1618512.}
}

\allowdisplaybreaks
\date{}
\maketitle

\begin{abstract}
Locally repairable codes (LRC) have recently been a subject of intense research due to theoretical appeal and
their application in distributed storage systems. In an LRC, any coordinate of a codeword can be recovered by accessing only few other 
coordinates. For LRCs over small alphabet (such as binary), the optimal rate-distance trade-off is unknown.
In this paper we provide the tightest known upper bound on the rate of linear LRCs of a given relative distance, an improvement over any previous 
result, in particular \cite{cadambe2013upper}.
\end{abstract}

\section{Introduction} 
\label{sec:introduction}

	Let $\cC(n,d) \subseteq \{0,1\}^n$ denote a binary linear code of length $n$ with minimum distance $d_{\min} =d$ and $\cC^\perp$ denote the dual of the linear code $\cC$.  
	A  code $\cC(n,d)$ can recover any codeword after the erasure of at most $(d-1)$ coordinate symbol. But, it requires knowledge of all other $n-d+1$ coordinates of the codeword for recovery. For certain scenarios such as distributed storage systems, in addition to minimum distance, codes with a small {\em{locality}} are desired. In such systems, to repair a single symbol of a codeword of $\cC$, we want to utilize information about few other codeword symbols. Locally Repairable Codes (LRC) \cite{gopalan2012locality}, with locality parameter $r$, allow the repair of any one symbol of a codeword by knowing at most $r$ other codeword symbols. Let $S$ be the subset of coordinates (of size at most $r$) required to repair a symbol at coordinate $i$ in the LRC. Then the set $R(i) \defeq S\cup \set{i}$ is called the repair group of coordinate $i$. We denote the linear LRC of length $n$, minimum distance $d$ and locality $r$ as $\cC(n,d,r)$.  

The dimension of $\cC(n,d)$ is defined as $k(\cC) = \log_2|\cC|$, and the rate as $R(\cC) = \frac{k}{n}$.
For a code $\cC(n,d,r)$, its dimension is known to be upper bounded by,
	\begin{equation}\label{eq:gopalan}
	k(\cC)\leq m-\floor*{\frac{m}{r+1}}\end{equation}
	where $m=n-d+1$ \cite{gopalan2012locality}. This bound is known to be achievable when the code is over  alphabet size at least $n$ \cite{tamo2014family}, but
	it is far from being tight for smaller, in particular binary, alphabets \cite{cadambe2013upper}. 
For binary code $\cC(n,d,r)$,  an upper  bound on the dimension of LRC was presented in  \cite{cadambe2013upper}:
\begin{equation}
k \le \min_{ t \in \integers_{+}} \Big[ tr + k_{\rm opt}(n  - t(r+1),d) \Big],
\end{equation}
where $k_{\rm opt}(n,d)$ is the optimum dimension of an error-correcting code of length $n$ and distance $d$.
	
	This bound can be translated into an
asymptotic bound on rate and relative distance, assuming $k/n \to R$ and $d/n \to \delta$ as $n \to \infty$:
\begin{eqnarray} R \leq \min_{0 \leq x \leq r/(r+1)} x + &\left(1-x(1+1/{r})\right)\nonumber\\
& \cdot R_{\rm opt}\left(\frac{\delta}{1-x(1+1/r)}\right) \label{eq:asymptoticonverse}\end{eqnarray}
where,
$R_{\rm opt}(\delta) \equiv  \lim_{n \to \infty} \frac{k_{\rm opt}(n,\delta n)}{n} $. The best known upper bound on the
rate of a locally repairable code is found when $R_{\rm opt}(\delta)$ is bounded by $R_{\rm MRRW}(\delta)$ given below:
\begin{equation}\label{MRRW}
R_{\rm MRRW}(\delta) = \min_{0<\alpha\le 1-4\delta} 1+ \hat h(\alpha^2) - \hat h(\alpha^2
+4\delta\alpha+4\delta)\,,
\end{equation}
with $\hat h(x) =h(1/2 -1/2\sqrt{1-x})$ and $h(x) \equiv -x\log_2 x -(1-x) \log_2 (1-x)$ is the binary entropy function.

A Gilbert-Varshamov type achievability bound was also presented in \cite{cadambe2015bounds}. These best known upper and lower bounds are well-separated. Despite
intense research on LRCs, and in particular on rate-bounds (cf. \cite{sihuang2016combinatorial}),   no result asymptotically better than  \eqref{eq:asymptoticonverse} is known.
In this paper, we provide improved upper bounds on the rate of linear LRCs that are tighter than \eqref{eq:asymptoticonverse}, for a range of rate-relative distance trade-off.

Our algebraic combinatorics methods are inspired by a recent work \cite{iceland2015coset}, that provides an improved estimate of rate of a code with sparse parity-check matrix. 
A combination of techniques from \cite{cadambe2013upper} and \cite{iceland2015coset} leads to improvement over the bounds in \cite{cadambe2013upper} for linear LRCs with arbitrary repair groups of size at most $r$. We further improve over these bounds for disjoint repair groups. Our bounds apply for the asymptotic case and give an improvement over the best known bounds on linear LRC for small alphabets. For clarity, we state our result for binary alphabet. 
	

The paper is organized as follows. In Section~\ref{sec:main}, we 
provide the main theorems: bounds on the rate of linear LRCs in general, and for the special case of 
disjoint repair groups.
The main techniques of coset counting is presented in Section~\ref{sec:rate_upper_bounds}.
The proof of the main results are presented subsequently.

\subsubsection*{Notations}
	For integers $a,b$ the relation $a \mid b$ denotes that $a$ divides $b$. $[m,n] \defeq \Set{m,m+1,\ldots,n}$ and $[n] \defeq \Set{1,2,\ldots,n}$. $\displaystyle \sqcup_i A_i$ denotes the union of disjoint sets $A_i$. $H((p_1,p_2,\ldots,p_n)^T) \defeq \sum_{i} -p_1 \Log{p_i}$ denotes the entropy for the distribution $\set{p_i}_i$ while $h(p) \defeq -p\log{p}-(1-p)\Log{1-p}$ denotes the binary entropy. The logarithms $\Log(.)$ are taken to be base $2$ throughout the paper. For a vector $\bfv \in \F_2^n$, let $\bfv(i)$ denote the symbol at coordinate $i$ of $\bfv$. Denote $s(\bfv)$ as the set $\Set{i\in [n] \given \bfv(i) \ne 0}$ and $\wt(\bfv)\defeq\abs{s(\bfv)}$. Define $\bfe_i, i\in [n]$ as a vector in $\F_2^n$ such that $s(\bfv) = \Set{i}$.

\section{Rate upper bounds for LRC} 

\subsection{Main results}\label{sec:main}
Our first result for general linear LRC rate is presented below.
\begin{theorem}[Linear LRC] \label{thm:linear}
The rate of a binary linear LRC $\cC(n,\delta n, r)$ can be at most,
\begin{align*}
R(\cC) \leq  \min_{0\leq x\leq r/(r+1)} x + &(1-x(1+1/r)) \\
& \cdot R_1\br*{\frac{\delta}{1-x(1+1/r)},r},
\end{align*}
where,
$
R_1(\delta,r) = h(\rho) - c(r+1,\rho),
$
and
$
c(w,\rho)  \defeq \frac{\log_2{e}}{8w^2}\br*{\frac{\rho^w}{2}}^{w+1}.
$ 
\end{theorem}

This bound is an improvement on \eqref{eq:asymptoticonverse} with $R_{\rm opt}$ replaced by $R_{\rm MRRW}$ and $R_1$ respectively. 
For LRCs with disjoint repair groups we have an even tighter bound on the rate.
\begin{theorem}[Linear LRC with Disjoint Repair Groups]\label{thm:disjoint}
Let, 
  	\begin{equation}\label{def_beta_i}  		
  	\beta_i(x) \defeq \begin{cases}
  		\frac{{r+1\choose i}}{x^i} & i<t \\
  		\frac{{r+1\choose i}}{2^{r\mmod 2} x^i} & i=t
  	\end{cases}
  	\end{equation}
  	The asymptotic rate $R(\cC)$ for a binary linear LRC $\cC(n,\delta n,r)$ with disjoint repair groups satisfies,
	\begin{equation}\label{UB_rate}
		R(\cC) \leq  \rho \Log\br{x} + \frac{\log\br{\sum_{i=0}^{t}\beta_i(x)}}{r+1} \Big\vert_{x=\mu}
  	\end{equation} 
	where
	$		
	\mu =  1
	$
	when $\sum_i i \frac{\beta_i(x)}{\sum_i \beta_i(x)}\Big\vert_{x=1} \leq (r+1)\rho$ and
	\begin{equation}\label{def_mu} 
\mu =   Roots^+((r+1) \rho \sum_{i=0}^t {r+1\choose i} x^{t-i}  - \sum_{i=0}^t {r+1\choose i} i\; x^{t-i})
 \end{equation}
  otherwise and
  	where $t \defeq \floor*{\frac{r+1}{2}}$ and $Roots^+(f(x))$ denotes the unique positive root of $f(x)=0$.
\end{theorem}

\begin{figure}
	\center
	\includegraphics[width=15cm,keepaspectratio]{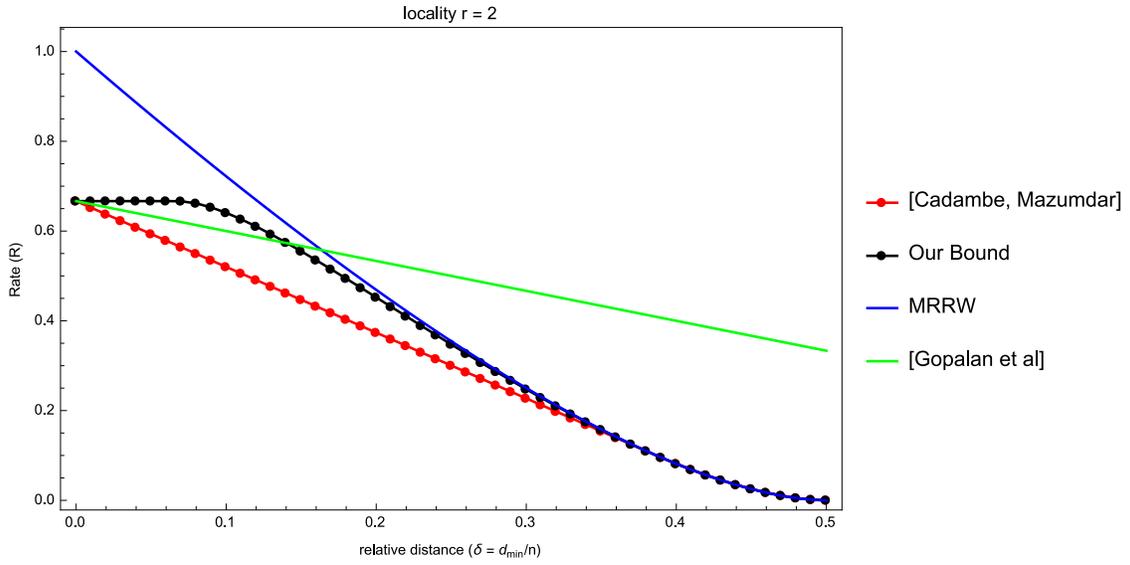}
	\caption{Comparison of the asymptotic upper-bound on rate (vs relative minimum distance) in \cref{UB_rate} with the existing bounds in \cite{gopalan2012locality,cadambe2013upper,mceliece1977new}.}\label{plot_r2}
	\end{figure}

\begin{figure}
	\center
	\includegraphics[width=13cm,keepaspectratio]{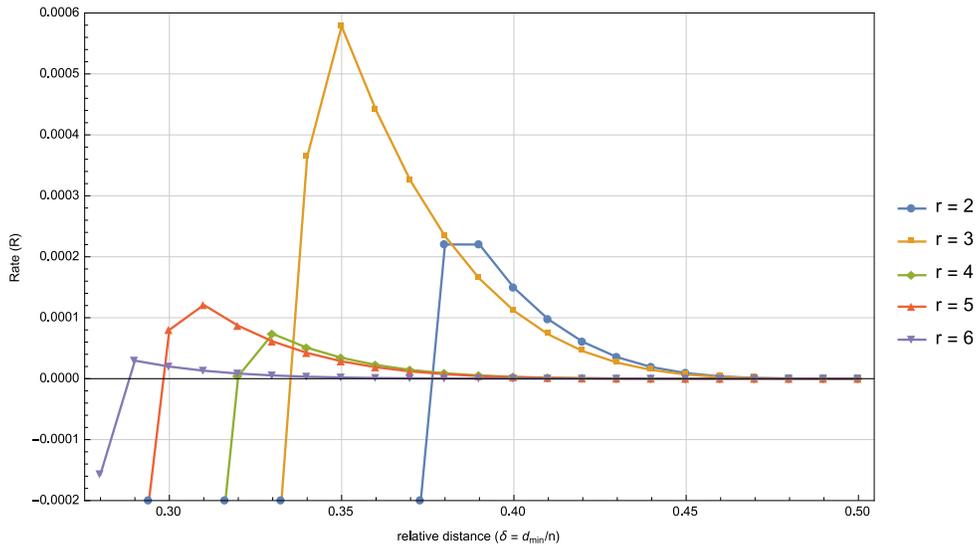}
	\caption{Difference between the bound in \eqref{eq:asymptoticonverse} (cf. \cite{cadambe2013upper}) and our bound \cref{UB_rate}.}\label{comparison_table}
	\end{figure}

In \cref{plot_r2} we compare the result of \cref{thm:disjoint} with the existing asymptotic bounds. The bounds converge together as $\delta$ and $r$ increase. But we can see from \cref{comparison_table} that our bound improves over the bound in \eqref{eq:asymptoticonverse} (cf. \cite{cadambe2013upper}) for $\delta \geq 0.38$ for $r=2$ and this range improves as $r$ increases.	
	
\subsection{The method of counting cosets}
\label{sec:rate_upper_bounds}
	
	The first MRRW bound provides the tightest (except for the second MRRW bound) upper bound on the rate of a code $\cC(n,\delta)$:
	\begin{equation}\label{general_MRRW_bound}
		R(\cC) \leq h(\rho), 
	\end{equation}
	where $\rho \defeq \frac{1}{2}-\sqrt{\delta\br*{1-\delta}}$. The bound is derived in \cite{mceliece1977new} using linear constraints on the  weight (distance)  distribution of the code $\cC$. Another way to derive the first MRRW bound for linear codes is to analyze the distance between the codewords in the {\em coset leader graph} of $\cC$\cite{friedman2005generalized}. These graphs are defined as follows \cite{friedman2005generalized}.

	{\definition The {\em coset leader graph} $\bb{T}(\cC) = (\cV,\cE)$ of a binary linear code $\cC$ is a graph with vertex set, $\cV = \Set*{\bf{x} + \cC^\perp \given \bfx \in \F_2^n }$ as the cosets of $\cC^\perp$. Two vertices in $\bb{T}(\cC)$ are connected iff the corresponding cosets are Hamming distance $1$ apart i.e. $\set{\bf{x}+\cC^\perp,\bf{y}+\cC^\perp} \in \cE$ $\implies \bfx = \bfy + \bfe_i + \bfv,$ for some $ \bfv \in \cC^\perp$ and $i\in [n]$.}

	It is known \cite{delorme1991diameter} that the eigenvalues $\lambda_1 \geq \lambda_2 \geq ...$ of the adjacency matrix of $\bb{T}(\cC)$ satisfy,
	$$\lambda_i = n-2 d_i$$
	where $d_1 \leq d_2 \leq ...$ are the weights of the codewords in $\cC$. Since $d_1 = 0$ for any linear code, we have $\delta = \frac{1}{2}(1- \lambda_2/n),$ i.e., 
the minimum distance of the code corresponds to the second eigenvalue of the coset leader graph. 	

	Note that, the shortest path length between the cosets $\cC^\perp$ and $\bfx+\cC^\perp$ is equal to the weight of the coset leader of $\bfx+\cC^\perp$. Denote the distance between two vertices in the graph $\bb{T}(\cC)$ as the length of the shortest path between them. Let $\Cl({v,r})$ denote the vertices in $\bb{T}(\cC)$ inside a ball of radius $r$ and centered at vertex $v$. Since, the graph $\bb{T}(\cC)$ is vertex transitive, the volume in the $\Cl({v,r})$ does not depend on $v$ and is written as $\Cl({r})$. Thus, $\Cl({r})$ also denotes the number of coset leaders of $\cC^\perp$ of weight not greater than $r$. From \cite[Theorem 1.1]{iceland2015coset}, we know that the asymptotic rate, $R(\cC)$ of code $\cC$ with relative minimum distance $\delta$, can be upper-bounded using the coset leader graph $\bb{T}(\cC)$ as,
	\begin{equation}\label{coset_leader_graph_ub}
		\abs{\cC} \leq 2^{o(n)} \abs*{\Clo},
	\end{equation}
	where, recall, $\rho \defeq \frac{1}{2}-\sqrt{\delta\br*{1-\delta}}$.
	Thus, the asymptotic MRRW bound on $R(\cC)$ in \cref{general_MRRW_bound} can be derived using a very general bound, 
	\begin{equation}\label{counting_cosets_simple}
		\Clo \leq \sum_{i\leq \rho n} {n \choose i}.
	\end{equation}

	Note that the binary linear LRC $\cC(n,\delta n, r)$ also satisfies \cref{coset_leader_graph_ub}. To improve the rate upper-bound \cref{general_MRRW_bound} for these codes, we can use their locality property to improve the counting upper-bound on $\Clo$, \cref{counting_cosets_simple}.

\subsection{Upper bound for linear LRCs: proof of \cref{thm:linear}} 
\label{subsec:upper_bound_on_lrc_rate}

	Consider a binary linear code $\cC$ with parity check matrix of constant row weights $w$. Then, for this code $\cC$ the following bound holds \cite[Thm 1.2]{iceland2015coset},
	\begin{equation}\label{samorodnitsky}
		\abs{\Clo} \leq 2^{-c(w,\rho) n} \abs{B(\rho n)}
	\end{equation}
	$B(t)$ denotes a ball of radius $t$ in $\F_2^n$. We describe a sketch of the proof of this bound from \cite{iceland2015coset} below. Our proof of \cref{thm:linear} will then follow.

	Consider a random binary vector $\bfv$ such that $\bfv(i) \sim Bernoulli(\rho)$ are  iid for $i\in [n]$. Then the probability $p\br{\rho}$ that the vector $\bfv$ is a coset leader of $\cC^\perp$ satisfies \cite[Lemma 2.1]{iceland2015coset},
	\begin{equation}\label{iceland1}
		p(\rho) \geq \Omega\br*{\frac{1}{\sqrt{n}}} \frac{\abs{\Clo}}{\abs{B(\rho n)}}.
	\end{equation}
	We now upper bound the probability $p\br{\rho}$ 
by determining a necessary condition for each coset leader. Next, we use \cref{iceland1} and the upper bound on $p(\rho)$ to find an upper bound on $\Clo$.

	Let the parity check matrix for the  code $\cC$ be $H = [\bfv_1 \bfv_2 \ldots \bfv_m]^T$ such that $\wt({\bfv_i}) = w, \forall i$. Then we divide the coordinates $[n]$ into partitions $\set{I_j}_{j\in [w]}$ using a recursive algorithm. Assume that $I_l, l>j$ have already been constructed. Then $I_j$ is constructed as in \cref{partition_samoro}. Note that each partition $I_j$ is such that $I_j = \bigsqcup_{l=1}^t U_{jl}$ for $U_{jl} : \abs{U_{jl}}=j$. Let the $\bfv_{jl}\in \Set{\bfv_i}_i$ be the vector containing the support of the $U_{ji}$. Then, it is easy to see that $s(\bfv_{ji})\setminus U_{ji} \subseteq  \bigcup_{i=1}^{j+1} I_i$.

	\begin{algorithm}
	\KwData{$\Set{\bfv_i}_{i\in [m]}$, $\Set{I_l}_{l=j+1}^w$}
	\KwResult{$I_j$}
	$I_j = \emptyset$, 
	$T = \Set{\bfv_i : i \in [m]}$ \;
	\While{$\exists \bfv_i \in T$}{
	\If{$\displaystyle \abs{s(\bfv_i)\setminus \bigcup_{l=j}^w I_l} = j$}{
		$\displaystyle I_j = I_j \cup \Set{s(\bfv_i)\setminus \bigcup_{l=j}^w I_l}$\;
	}
	$T=T\setminus \Set{\bfv_i}$
	}
	\caption{Recursive algorithm for partitioning the coordinates in the code $\cC$}
	\label{partition_samoro}
	\end{algorithm}

	An important property of the partitions $\set{I_j}_j$ is as follows \cite[Lemma 2.2]{iceland2015coset},
	{\lemma There exists a partition $I_k$ such that,
	$$\abs{I_k} \geq \max \Set*{ A \sum_{j>k} \abs{I_j}, B }$$
	where $A = \frac{2}{\rho^w}$ and $B = \frac{n}{2w A^w}.$
	\label{iceland_lemma22} }

	The proof of \cref{iceland_lemma22} follows from the requirement that there must exist at least one $k \in [n]$ such that $\abs{I_k}\geq n/w$. Let the coordinate $k$ be such that \cref{iceland_lemma22} is satisfied i.e. $\abs{I_k} \geq \max \Set*{ A \sum_{j>k} \abs{I_j}, B }$. Then, we claim that the support for every coset leader $\bfx$ can contain at most $\frac{\rho^k}{2} t$ of the $k$-tuples $\set{U_{kl}}_{l=1}^t$. This can be immediately  seen, since $$\bfy = \bfx + \sum_{i \in S}\bfv_{ki}$$ has weight less than $\bfx$ for the set $S \defeq \Set{i \given U_{ki} \subseteq s(\bfx)}$ such that $\abs{S}\geq \frac{\rho^k}{2}$. Thus, using this necessary condition for the coset leader $\bfx$ we can upper bound $p(\rho)$ in \cref{iceland1} as follows,
	\begin{eqnarray}
		p(\rho)\leq \prob\Big({\rm Binomial}(\rho^k,t)\leq & \frac{\rho^k}{2} t\Big) \leq \Exp\Big(-\frac{\rho^k\cdot t}{8}\Big)\nonumber\\
		& \stackrel{(a)}{\leq} 2^{-c(w,\rho) n} ,
	\end{eqnarray}
	where inequality $(a)$ above follows since for $k$ satisfying \cref{iceland_lemma22} we have,
	$${\rho^k} \cdot t \leq \frac{\rho^w}{w} B = \frac{\rho^w}{w}\cdot \frac{1}{2wA^w} \cdot n. $$
	
	Here, we note that the upper-bound \cref{samorodnitsky} 
	requires only the assumption that the parity check matrix $H$ of $\cC$ contains constant weight row vectors $\bfv_i$ whose support covers the coordinates $[n]$.
Note that for an LRC, every coordinate must have a repair group of size at most $r+1$. Hence in the parity check matrix, there must be rows of weight $r+1$ that
covers all the coordinates by their support.	
	 Thus, we have the following upper bound for  an LRC $\cC(n,\delta n,r)$.
	{\lemma \label{samoro_thm1}  The asymptotic rate $R(\cC)$ for a binary linear LRC $\cC(n,\delta n,r)$ satisfies,
	$$R(\cC) \leq \underbrace{h(\rho) - c(r+1,\rho)}_{R_1(\delta,r)}$$
	where and $c(w,\rho) \defeq \frac{\log_2{e}}{8w^2}\br*{\frac{\rho^w}{2}}^{w+1}$.
	}

	The upper bound in \cref{samoro_thm1} gives an upper-bound worse than the bound in \eqref{eq:asymptoticonverse}. 
	But we can construct a better bound by using the bound in \cref{samoro_thm1} inside the shortening bound in \eqref{eq:asymptoticonverse}. 
A mild change in the argument of 	
	\cite{cadambe2013upper} first leads to the following bound.
	
{\theorem The rate for an LRC code $\cC(n,\delta n,r)$ satisfies,
		\begin{align*}
		R(\cC) \leq \min_{0\leq x\leq r/(r+1)} x + &(1-x(1+1/r)) \\
		&\cdot R_{\rm opt}\br*{\frac{\delta}{1-x(1+1/r)},r}
		\end{align*}
	where $R_{\rm opt}(\delta,r)$ is the optimal asymptotic rate of $\cC(n,\delta n,r)$.
	}

	In \cite{cadambe2013upper},  the MRRW bound for $R_{\rm opt}(x,r)$ was used. If we use the bound in \cref{samoro_thm1} for $R_{\rm opt}(.)$ we have,
	\begin{align*}
	R(\cC) \leq  \min_{0\leq x\leq r/(r+1)} x + & (1-x(1+1/r)) \\
	& \cdot R_1\br*{\frac{\delta}{1-x(1+1/r)},r}
	\end{align*}
	where $R_1(.)$ is as defined in \cref{samoro_thm1}.

This completes the proof of \cref{thm:linear}.


\subsection{Upper bound on LRC for disjoint repair groups: proof of \cref{thm:disjoint}} 
\label{subsec:upper_bound_disjoint_repair_groups}

	In this section we consider the special case of LRC codes with disjoint repair groups. For this case too, the bound in \cref{samoro_thm1} applies. But we can improve that bound significantly in this situation. Consider a binary linear LRC $\cC(n,\delta n,r)$ with disjoint repair groups $R_j, j\in \brac*{\frac{n}{r+1}} $. For clarity, in this section we assume $(r+1) \mid n$. The parity check-matrix of $\cC$ must be of the form,
	$$
 		H = \begin{pmatrix}
	 			\bf{v}_1 \\
	 			\vdots \\
	 			\bf{v}_{n^\prime} \\
	 			H^\prime
	 		\end{pmatrix}
	 $$
 	where $n^\prime = \frac{n}{r+1}$, $s(\bf{v}_i) \cap s(\bf{v}_j) = \emptyset$ and $R_i \defeq s(\bf{v}_i)$ are such that $\abs{R_i} = r+1$ (even if the repair groups are of smaller size, we can add redundant coordinates in them to equal $r+1$). Then, $\bfv \in \Clo$ satisfies the following constraint,
 	\begin{subequations}\label{constraint1}
 	\begin{gather}
 		\wt\br*{\bf{v}} \leq \min_{S\subseteq [n^\prime]} \wt\br*{\bf{v}+\sum_{i\in S} \bf{v}_i}  \label{constraint1_minwt}\\
 		\wt\br*{\bf{v}} \leq \rho n  \label{constraint1_maxwt}\\
 		\abs*{\Clo \cap (\bfx+\spn\br*{\Set{\bfv_i}_i})} \leq 1 \;\forall \bfx\in \F_2^n. \label{constraint1_uniqueness}
 	\end{gather}
 	\end{subequations}
 	We use \cref{constraint1} to find a better upper bound on $R(\cC)$.

  	\begin{proof}[Proof of \cref{thm:disjoint}] 	
  	We find all vectors satisfying \cref{constraint1} to upper bound $\abs{\Clo}$ and to subsequently get an upper bound on $R(\cC) = \Log{\abs{\Clo}}/n$. Using \cref{constraint1_maxwt,constraint1_minwt} we must have,
	\begin{equation}\label{UB_CLset}
		\Clo \subseteq \underbrace{\Set*{\bf{x}\in \F_2^n \given \wt{\bf{x}} \leq \rho n, \; \abs{s{\br{\bfx}}\cap R_i} \leq \s\;\forall R_i}}_{\cS} .
	\end{equation}
	The number of words in  $\cS$ is given by \begin{equation}\label{generating_fn}
		\sum_{k\leq \rho n} [x^k] \br*{1+\sum_{i=0}^{t}{r+1 \choose i} x^i}^{\frac{n}{r+1}}
	\end{equation}
 	where $t=\s$ and $[x^k] g(x)$ denotes the coefficient of $x^k$ in $g(x)$.	  

  	We can further reduce the bound in \cref{UB_CLset} using \cref{constraint1_uniqueness}. Since, $\bfv$ and $\bfv+\bfv_i$ are of equal weight and are in the same coset we need to count only one of them in $\cS$. Thus, for $r+1 = \mbox{even}$, we can improve the bound in \cref{generating_fn} as follows,
	\begin{equation}\label{UB_CLset1}
		\abs{\Clo} \leq \sum_{k\leq \rho n} [x^k] \underbrace{\br*{1+\sum_{i=0}^{t-1}{r+1 \choose i} x^i+ \frac{{r+1 \choose t}}{2^{r\mmod 2}} x^t}^{\frac{n}{r+1}}}_{g(x)}.
	\end{equation}
	We now see that the upper bound in \cref{UB_CLset1} is dominated (asymptotically) by $\max_{k} [x^k] g(x)$ i.e. 
	$$\frac{\Log{\displaystyle\sum_{k\leq \rho n} [x^k] g(x)}}{n} = \max_{k\leq\rho n}\frac{\Log{[x^k] g(x)}}{n},$$
	as $n \to \infty$.
	Thus we have,
	\begin{align}\label{UB_CL_concave}
		R(\cC)&\leq \max_{k\leq\rho n}\frac{\Log{[x^k] g(x)}}{n} \nonumber \\
			  &= \max_{\substack{\sum_{i}\alpha_i = \frac{1}{r+1}\\ \sum_{i}i \alpha_i \leq \rho } } \frac{1}{r+1} H\br{(r+1)  \cdot \vect[0]\alpha[t]} \nonumber\\
			  &+ \sum_{i=0}^{t-1} \alpha_i \Log{{r+1 \choose i}} +  \alpha_t \Log{\frac{{r+1 \choose t}}{2^{r\mmod 2}}}.
  	\end{align}

  	Since, \cref{UB_CL_concave} represents a concave optimization  with linear constraints, we can find the optima in \cref{UB_CL_concave} analytically. The optimum value is achieved at $\alpha_i = \frac{\beta_i(x)}{(r+1)\sum_{i=1}^t \beta_i(x)}\big\vert_{x=\mu}$ for $\beta_i(x)$ and $\mu$ as defined in \cref{def_beta_i,def_mu} and \cref{UB_CL_concave} evaluates to \cref{UB_rate}.
  	\end{proof}



\bibliographystyle{abbrv}
\bibliography{references}

\end{document}